\documentclass[12pt]{article}
\usepackage{graphicx,epsfig,natbib,multicol}
\usepackage{hyperref}
\usepackage{wrapfig}
\usepackage{times}
\usepackage{subfigure}
\usepackage{simplemargins}
\setleftmargin{0.9in}
	\setrightmargin{0.9in}
	\settopmargin{0.9in}
	\setbottommargin{0.9in}
\usepackage{macros_rudnick}

\usepackage[compact]{titlesec}


\bibpunct[;]{(}{)}{;}{a}{}{,}

\begin{document}
%\maketitle
%\noindent\HRule

%\renewcommand{\thepage}{D--\arabic{page}}

%\everypar{\looseness=-1}

%\hline
%\begin{center}
%Project description
%\end{center}
%\hline

%\begin{multicols}{2}

%\HRule
%\vspace{-1.5in}
\begin{center}{\LARGE Roman CCS White Paper}\end{center}
\vspace{0.5cm}

\noindent \textbf{\Large Roman--Cosmic Noon: A Legacy Spectroscopic Survey of Massive}

\smallskip
\noindent \textbf{\Large Field and Proto-cluster Galaxies at $2 < z < 3$}

\noindent \textbf{Roman Core Community Survey:} High Latitude Wide Area Survey

\noindent \textbf{Scientific Categories:} galaxies; large scale structure of the universe

\noindent \textbf{Additional scientific keywords:} Galaxy environments, Galaxy evolution, Galaxy formation, High-redshift galaxies, Quenched galaxies, Extragalactic Legacy And Deep Fields, Galaxy clusters, Galaxy groups

\noindent\textbf{Submitting Author:} Gregory Rudnick,University of Kansas, grudnick@ku.edu

\noindent \textbf{List of contributing authors:}
Yannick Bah\'e, EPFL, Switzerland, yannick.bahe@epfl.ch,\\
Michael Balogh, University of Waterloo, Canada, mbalogh@uwaterloo.ca, \\
Mike Cooper, UC Irvine, cooper@uci.edu,  \\
Nina Hatch, Nottingham University, UK, Nina.Hatch@nottingham.ac.uk, \\
Benedetta Vulcani, Padova Observatory, Italy, benedetta.vulcani@inaf.it, \\
Gillian Wilson, UC Merced, gwilson@ucmerced.edu,\\ 
Gianluca Castignani, University of Bologna, Italy, gianluca.castignani@unibo.it, \\
Pierluigi Cerulo, Universidad de Concepci\'on, Chile, pierluigi.cerulo@gmail.com, \\
Gabriella De Lucia, Astronomical Observatory of Trieste, gabriella.delucia@inaf.it, \\
Ricardo DeMarco, Universidad de Concepci\'on; rdemarco@astro-udec.cl, \\
Benjamin Forrest, UC Davis, bforrest@ucdavis.edu,  \\
Pascale Jablonka, EPFL, Switzerland, Pascale.Jablonka@epfl.ch

\noindent \textbf{Abstract}: Protoclusters are the densest regions in the distant universe ($z>2$) and are the progenitors of massive galaxy clusters ($M_{halo}>10^{14}{\rm M}_\odot$) in the local universe. They undoubtedly play a key role in early massive galaxy evolution and they may host the earliest sites of galaxy quenching or even induce extreme states of star formation. Studying protoclusters therefore not only gives us a window into distant galaxy formation but also provides an important link in our understanding of how dense structures grow over time and modify the galaxies within them.  Current protocluster samples are completely unable to address these points because they are small and selected in a heterogeneous way.  We propose the Roman-Cosmic Noon survey, whose centerpiece is an extremely deep (30ksec) and wide area (10 deg$^2$) prism slitless spectroscopy survey to identify the full range of galaxy structures at $2<z<3$.  This survey will include 1500 uniformly selected protoclusters, their surrounding cosmic web environments, and at least 15,000 protocluster galaxies with $M_\star>10^{10.5} {\rm M}_\odot$ across the full range of star formation histories as well as many more lower mass star-forming galaxies.  The survey will also contain field galaxies to much lower masses than in the High Latitude Wide Area Survey, but over an area dwarfing any current or planned deep spectroscopy probe at $z>2$.  With the prism spectroscopy and some modest additional imaging this survey will measure precise stellar mass functions, quenched fractions, galaxy and protocluster morphologies, stellar ages, emission-line based SFRs, and metallicities.  It will have extensive legacy value well beyond the key protocluster science goals.

\clearpage

\begin{center}
\Large{Roman--Cosmic Noon: A Legacy Spectroscopic Survey of Massive Field and Proto-cluster Galaxies at $2 < z < 3$}\\
\end{center}
\bigskip

Numerous studies over the last few decades have allowed us insights into how galaxy evolution depends on environment at $z>1$. Current evidence suggests that even when the Universe was only a few billion years old, cluster galaxies were already more evolved than their field counterparts (Nantais et al. 2017; Lee Brown et al. 2017), but we do not know when these differences developed and whether they began prior to, or upon infall into clusters (van der Burg et al. 2020; Webb et al. 2020).  These speak to some of the longest-standing questions in galaxy evolution.  Protoclusters are the place to answer these questions, as they provide a wide range of local densities over which to explore galaxies that will end up in clusters at lower redshift (Chiang et al. 2017; Muldrew et al. 2018).  However, environmental studies at $z>2$ have only scratched the surface of this question and are plagued by small and heterogeneous protocluster samples with a high degree of variance between individual systems.  For example, protoclusters at $z>2$ have been selected using X-rays for the most massive sources (Gobat et al. 2011), in the rest-frame Ultraviolet by overdensities of unobscured star-forming galaxies using Ly-$\alpha$ emitters or Lyman Break Galaxies (Steidel et al. 2005; Lee et al. 2014; Toshikawa et al. 2016; Shi et al. 2021), in the rest-frame optical by photometric redshifts (Spitler et al. 2012) or deep observed NIR spectroscopy (Kubo et al. 2021; McConachie et al. 2022), as projected overdensities around radio galaxies (Hatch et al. 2011), or as overdensities of dusty star-forming galaxies detected by their FIR emission (Casey et al. 2015).  Each of these selection techniques carries its own biases both in regard to galaxy evolution, as manifested by the types of galaxies (passive vs. star-forming), and in regards to the types of protoclusters (dense vs. extended).  In addition, given the relatively modest galaxy overdensities in these systems it is difficult to conclusively establish membership without deep spectroscopy.  The dearth of large samples of homogeneously selected clusters with spectroscopy makes it impossible to address the following pressing problems: What role do protoclusters play in quenching galaxies or boosting star formation at $z>2$?  If they do play a role, does it happen in protocluster cores or in the less dense extended structures and with what efficiency?  What is the variance in protocluster properties and how does this connect to the various paths of galaxy evolution?  How is the Baryon Cycle modified in protocluster galaxies, as probed through the interplay between star formation, gas content, and metallicity as well as through outflows?  

\begin{figure}[t!]
\centering
\includegraphics[width=0.95\textwidth]{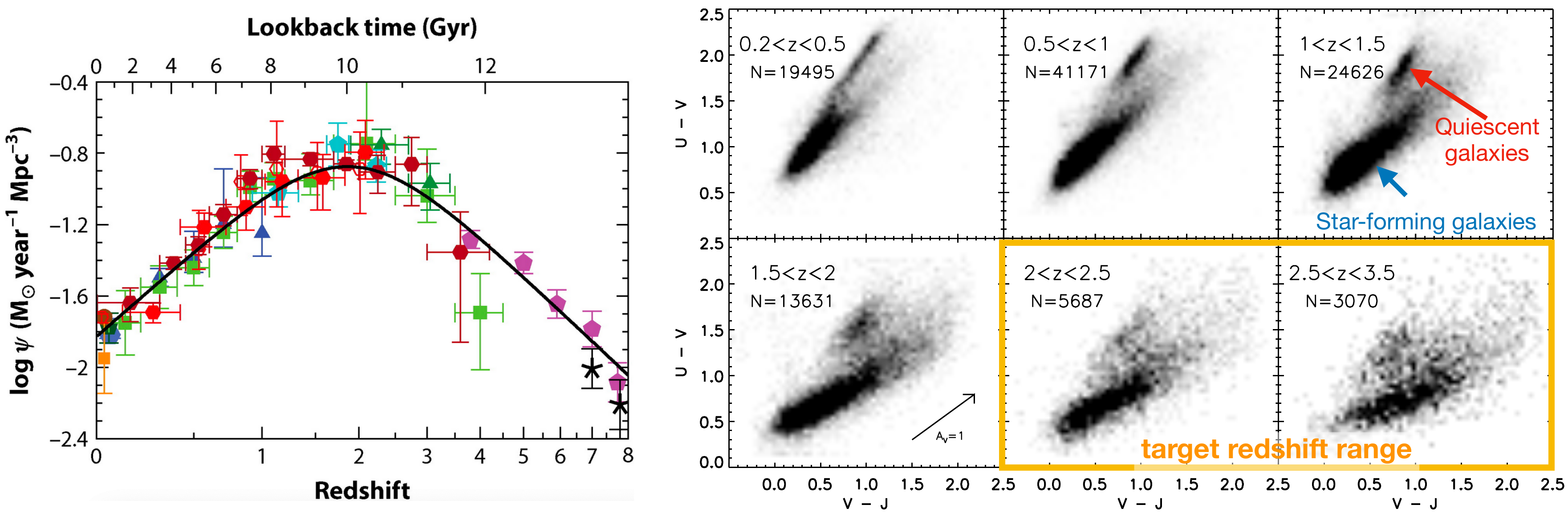}
\vspace*{-0.1in}
\caption{{\bf Left:} Star formation rate history of the Universe showing a peak at $2< z< 3$ (Madau \& Dickinson 2014).  {\bf Right:} $UVJ$ diagrams as a function of redshift from the COSMOS UltraVISTA DR3 catalog
%\citep{muzzin13b}.
(Adam Muzzin, private communication).  Our target redshift range is indicated by the golden box.
As \textbf{Fig.~\ref{fig:quenching-epoch}} also shows, 
``cosmic noon" ($2 < z< 3$) is the epoch at which significant numbers of massive galaxies first begin to transition from star-forming to quiescent.  Our proposed survey will enable,  for the first time at this critical epoch, the determination of \textit{spectroscopic redshifts} and ages for a mass-complete sample of quiescent galaxies  for $\sim 1500$ protoclusters.
}
\label{fig:UVJ}
%\vspace{-0.2in}
\end{figure}

\begin{figure}[t]
\centering
\includegraphics[width=0.98\textwidth]{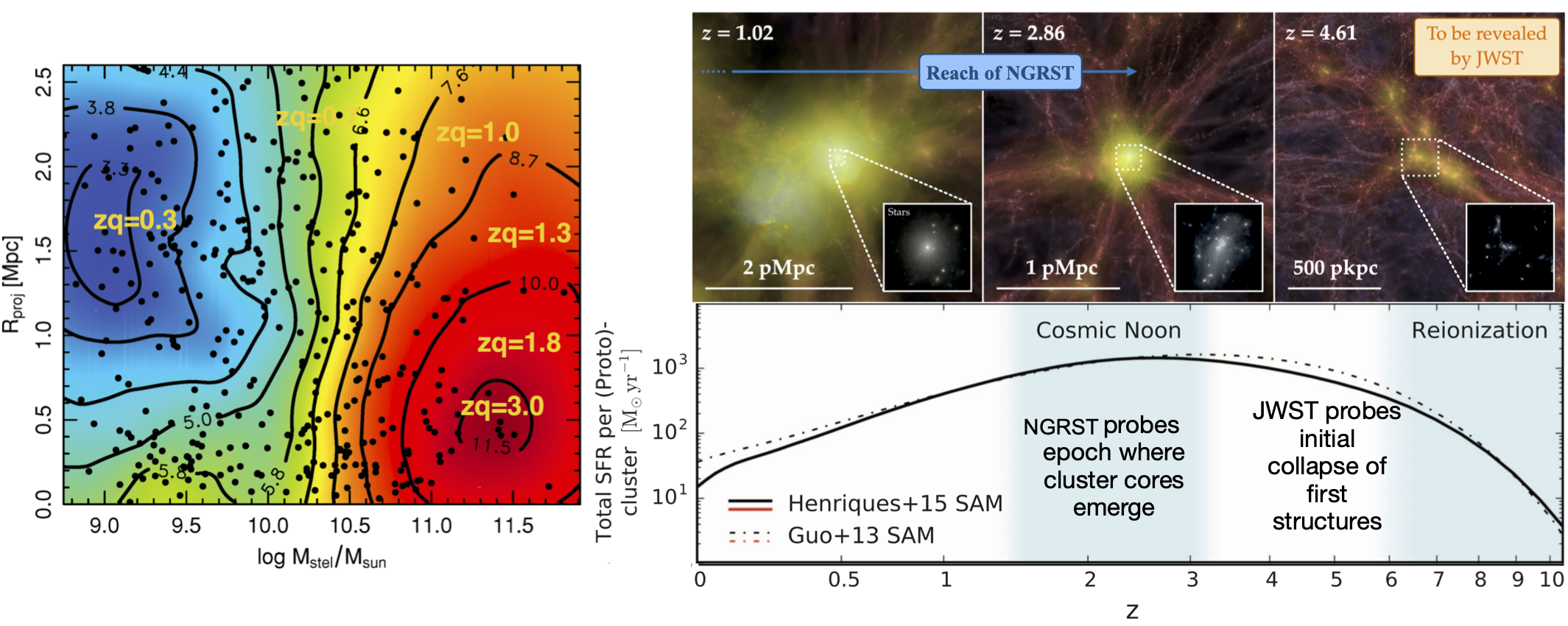}
\caption{ {\bf Cosmic Noon is the Epoch of Quenching in Protocluster Cores - Left:} Analysis of the spectra of quiescent members in the Coma cluster shows that $2 < z< 3$ is the epoch of quenching in protoclusters. 
``zq" is the redshift at which  each galaxy must quench in order to satisfy the age contours shown in black (Smith et al. 2012). {\bf Right:} Hydrodynamical simulations  (upper; Bah\'e et al. 2017) and semi-analytic models (lower; adapted from Chiang et al. 2017) of representative protoclusters
%(\citealt{henriques15, guo13})  
support the observational Coma cluster age constraints, showing that protocluster cores form and quenching begins in those cores in earnest during the ``cosmic noon" epoch at  $2 < z< 3$.  In the upper right panel, the large frames show the gas while the insets show the stars.  NGRST can probe this critical epoch, when the ICM first becomes prevalent and when the stellar cores of clusters first emerge.
%}
%\vspace*{0.4in}
}
\label{fig:quenching-epoch}
%\vspace{-0.5in}
\end{figure}

\begin{figure}[t!]
\vspace{-0.2cm}
\centering
\centering
\includegraphics[width=0.8\textwidth]{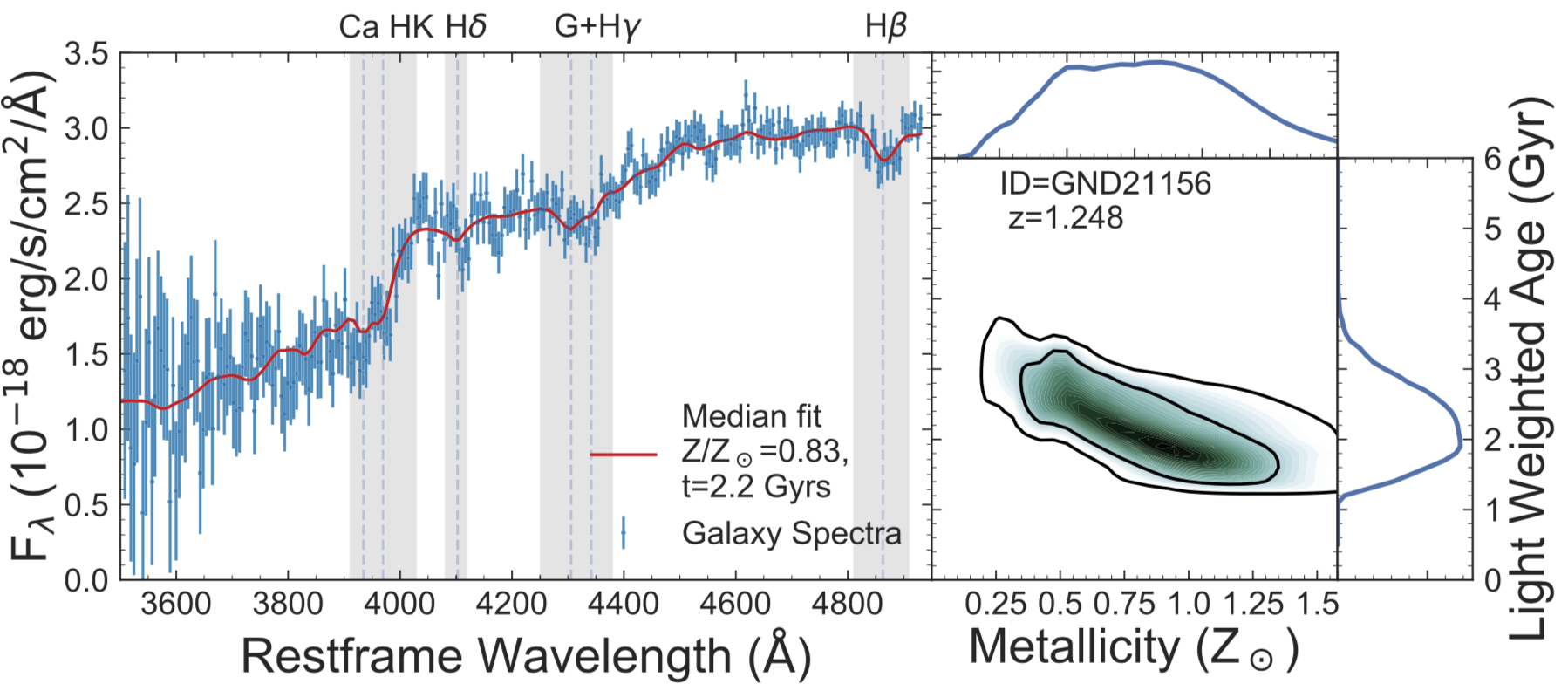}
\caption{Proof-of-concept demonstration showing a spectrum from the 12-orbit depth G102/F105W CLEAR survey which used the G102L WFC3 grism to target quiescent galaxies at redshift $1 < z< 1.8$ (Estrada-Carpenter et al. 2019). 
The P127 observations in our proposed program have similar rest-frame wavelength and stellar mass limits to CLEAR, but at $2 < z< 3$.  Combined with the exquisite photometry in these fields, our spectroscopic observations will yield 1~Gyr uncertainties on the ages for individual galaxies after marginalizing over the metallicity.
}
%\vspace{-0.5cm}
\label{fig:examples}
\end{figure}

Key to addressing these topics are a large sample of protoclusters with robust spectroscopic membership and deep optical-NIR photometry.  With the right observing strategy, the Nancy Grace Roman Space Telescope (NGRST) has the potential to revolutionize our understanding of the earliest dense regions in the Universe.  We therefore propose a deep ($\sim 30$ ksec) spectroscopic survey within an LSST Deep Drilling Field (DDF) with the P127 prism and with supplementary imaging in the F148 and F213 filters over 10~deg$^2$.  We target $2<z<3$ (``Cosmic Noon") as this redshift range is the first epoch where quenched galaxies appear in large quantities (Fig.~\ref{fig:UVJ}), where protocluster cores form (Fig.~\ref{fig:quenching-epoch}) and where the 4000\AA\ break is readily visible with the P127 prism, which enables robust redshifts and age constraints (Fig.~\ref{fig:examples}).

\begin{center} \textbf{Science requirements and Survey Design}
\end{center}
\textbf{Protocluster sample and area:} The protocluster discovery space of NGRST will be maximized with a survey that probes: \textbf{1)} a large enough volume to find substantial numbers of protoclusters, \textbf{2)} that is deep enough to detect galaxies below $L_\star$ regardless of their star formation history, and \textbf{3)} which has sufficiently deep spectroscopy and multi-band imaging to identify and characterize members. Specifically, we require a large enough sample of protoclusters to fully characterize the diversity of properties including: spatial concentration, morphology (core dominated vs.~filamentary), richness (or mass), and galaxy composition.  A key science goal is to reveal how protoclusters evolve over the $2<z<3$ range.  With 2-3 bins in each of these 4 properties, 2 bins in redshift, and 10 protoclusters in each bin we will require $\sim 1500$ protoclusters.  We estimate the required area to observe this number of protoclusters using the number density of the main progenitors of $z=0$ halos with $M>10^{14}$~M$_\odot$, which is $9.25\times 10^{-6}$~cMpc$^{-3}$ (Schaye et al. 2023).  The volume needed to obtain 1500 cluster progenitors (protoclusters) at $2<z<3$ therefore corresponds to a $300\times 300$~cMpc area, or $\sim 10$~deg$^2$.  

\textbf{Galaxy mass limit and galaxy sample size:} We wish to obtain unambiguous spectroscopic redshifts for galaxies of all star formation histories (SFHs).  This in turn will allow us to confirm protoclusters regardless of their galaxy composition.  This is critical as we know from surveys like MAGAZ3NE (McConachie et al. 2022) that there are many quenched galaxies in the distant Universe.  For galaxies without emission lines we will measure redshifts using the 4000\AA\ Break.  A survey based only on emission-line redshifts will potentially miss the protoclusters that host the highest fraction of quenched galaxies. Likewise, we need to go faint enough to both obtain enough galaxy spectra to unambiguously identify protoclusters and to characterize the massive galaxy population in the protoclusters.  As demonstrated in Lee-Brown et al. (2017), Cooke et al. (2016), and Newman et al. (2014), the passive fraction in three $z\sim 1.6$ protoclusters starts to exceed that in the field at log($M_{star}/$M$_\odot)>10.5$.  To probe the effect that protoclusters may have on massive galaxies we therefore need to reach at least this mass limit.  Using the COSMOS2020 catalog (Weaver et al. 2022) we have found that quiescent galaxies at this stellar mass at $z=2.7$ have H(AB)=24.6 and use this for our benchmark depth.

Down to this stellar mass limit Edward et al. (in prep) finds that there are $\sim 10$ galaxies per protocluster within 1.5~Mpc of the protocluster core.  This will yield a protocluster galaxy sample of 15,000 protocluster galaxies.  In addition, there will be many more field galaxies projected along the line of sight, but for which this program will obtain redshifts and SED constraints.

\textbf{Imaging survey depth and integration time:} For our stellar mass limit for quiescent galaxies (log($M_{star}/$M$_\odot)>10.5$) we wish to measure structural parameters of galaxies, which requires an integrated $S/N=20$ in an $r=0.66$ arcsec (6 pixel) aperture (van der Wel et al., 2012) for an object of F158(AB)$ = 24.6$.  This $S/N$ must be attainable for galaxies of the expected spatial profiles of our targets,  which have $r_e\sim 2.25$ kpc, corresponding to $r\sim 0.3$ arcsec at $z\sim 2.5 - 3.0$ (van der Wel et al., 2014).  We perform all our calculations assuming $1.44\times$ the minimum zodiacal light.  Using the online sensitivity tables \footnote{\href{https://roman.gsfc.nasa.gov/science/anticipated\_performance\_tables.html}{https://roman.gsfc.nasa.gov/science/anticipated\_performance\_tables.html}} we find that we can achieve this $S/N$ in 2460s of integration per pointing in F158 and 4096s in F184.  While F213 imaging would be very useful for its ability to access the rest-frame $V$-band, the exposure times are significantly longer, at 18ksec per pointing.  It is, however, possible to relax the $S/N$ constraints for F213 to $S/N=10$, which will allow us to detect galaxies but not robustly measure the sizes for the faintest objects.  This corresponds to an exposure time of 4500s.

\textbf{Spectroscopic survey depth and integration time:} For our spectroscopic component we determined the necessary integration time by scaling from the results of Lee-Brown et al. (2017) and Balogh et al. (2021), who showed that quiescent galaxies at $1<z<2$ can have their redshifts measured with $90\%$ success at $S/N\sim 7$ per resolution element at rest-frame wavelengths of $0.45$\micron.  This corresponds to a $S/N\sim 5$ per pixel for the Nyquist sampled P127 spectral element.  We use the online sensitivity calculator to estimate that we will need $\sim 30$ksec integration time to reach our $S/N$ requirement.   At this depth the survey will also be extremely effective at detecting emission line sources to much lower masses and measuring precise redshifts.  This will not only provide an indispensable characterization of the cosmic web, but also will allow for the use of [OII]-based SFR indicators.

\textbf{Instrument modes:} Our proposed survey will be well-served by the F106, F129, F158 filters to be used as part of the High Latitude Wide Area (HLWA) Survey.  For our science, going to as long wavelength as possible is the most important so as to better mimic a stellar-mass selection.  We would therefore advocate either replacing F184 with F213 at comparable depth, or even better adding F213 at comparable depth.  Adding additional deep imaging with F146 will also allow the photometric coverage to better sample the 4000\AA\ Break at our galaxy redshifts.

We request the P127 grism because its wider wavelength range and higher throughput will maximize the number of redshifts down to our survey depth.  Experience with HST/WFC3 has shows that this depth is sufficient to obtain passive galaxy redshifts.  In dense protocluster cores, it is likely that galaxy spectra will overlap and this contamination can affect our redshift success rate.  We therefore propose that the spectroscopic observations be split into $5\times 6000$sec visits with as large of offsets in orientation as possible between visits.  Extensive HST experience in clusters (HST-GLASS; Treu et al 2015; Lee-Brown et al. 2017) has shown that this makes it possible to better deal with spectral contamination.

\textbf{Field Choice:}  The 20~deg$^2$ area of the deep High Latitude Survey (HLS) should be deep enough for our imaging purposes, though 1-2 extra filters may need to be added to this survey (see above).  If the HLS is done in an LSST Deep Drilling Field we will naturally obtain the necessary ultra-deep optical data needed for complete optical/NIR photometric coverage. The most significant addition of our program would be deep (30ksec) prism spectroscopy over this field, at different orientations.  The minimum area for our spectroscopy would be $\sim 10$~deg$^2$, though certainly covering larger areas would increase our statistical power.

\textbf{Total program time and scheduling:} It will take 35 pointings to cover 10~deg$^2$, or 24 hours for the imaging component and 300 hours for the spectroscopic component.  The total time needed is well within that allocated for the High Latitude Wide Angle (HLWA) Survey.  If conducted within the Continuous Viewing Zone (CVZ) the requirement of multiple roll angles would be straightforward to satisfy.

\textbf{The legacy value of this program for NGRST:} Thanks to the extremely high multiplexing of NGRST, this program will obtain spectra for every galaxy over a large area and with $\sim 10\times$ longer exposure times than the HLWA Survey. The long-term and broad value of such surveys has been convincingly demonstrated through the analogous 3DHST grism survey with HST (Nelson et al. 2012, Momcheva et al. 2016) and we expect a similarly transformative scientific return from a program like ours.  This science will not be possible with wide-field highly multiplexed spectrographs on large telescopes (e.g. MOONS/VLT, PFS/Subaru) that only can obtain redshifts for emission galaxies at modest mass and have gaps in redshift coverage because of atmospheric transparency gaps. The resultant sample of protoclusters will provide the benchmark sample for all protocluster follow-up with JWST, ALMA, and ELTs, for the foreseeable future and will yield invaluable insights about early galaxy evolution in the Universe?s densest environments.

\begin{center}
\textbf{References}
\end{center}
Bah\'e, Y. M. et al. 2017, MNRAS, 470, 4186\\
Casey, C. et al. 2015, APJ, 808, L33\\
Chiang, Y.-K. et al. 2017, ApJ, 844, L23\\
Cooke, E. A. et al. 2016, ApJ, 816, 83\\
Estrada-Carpenter, V. et al. 2019, ApJ, 870, 133\\
Gobat, R. et al. 2011, A\&A, 526, A133\\
Hatch, N. et al. 2011, MNRAS, 410, 1537\\
Kubo, M. et al. 2021, APJ, 919, 6\\
Lee-Brown, D. B. et al. 2017, ApJ, 844, 43\\
Lee, K.-S. et al. 2014, ApJ, 796, 126\\
Madau, P. et al. 2014, ARA\&A, 52, 415\\
McConachie, I. et al. 2022, ApJ, 926, 37\\
Momcheva, I. G. et al. 2016, ApJS, 225, 27\\
Muldrew et al. 2018, MNRAS, 473, 2335\\
Nantais , J. et al. 2017, MNRAS, 465, L104\\
Nelson, E. et al. 2012, ApJL, 747, L28\\
Newman, A. B., Ellis, R. S., Andreon, S., et al. 2014, ApJ, 788\\ 
Schaye, J. et al. 2023, submitted to MNRAS, arXiv:2306.04024\\
Shi, K. et al. 2021, ApJ, 911, 46\\
Shimakawa, R. et al. 2018, MNRAS, 473, 1977\\
Smith, R. J. et al. 2012, MNRAS, 419, 3167\\
Spitler, L. et al. 2012, ApJ, 748, L21\\
Steidel, C. et al. 2005, ApJ, 626, 44\\
Toshikawa et al. 2016, ApJ, 826, 114\\
Treu, T. et al. 2015, ApJ, 812, 114\\
van der Burg, R. F. J. et al. 2020, A\&A, 638, A112\\
van der Wel, A. et al. 2012, ApJS, 203, 24\\
?. 2014, ApJ, 788, 28\\
Weaver et al. 2022, ApJS, 258, 11\\
Webb, K. et al. 2020, MNRAS, 498, 5317\\
Yuan, T. et al. 2014, ApJ, 795, L20

%\setcounter{page}{1}
%\clearpage
%\bibliographystyle{/Users/grudnick/LaTeX/Bibtex/apj1lim}
%\bibliography{references,rfinn,references_2}

\end{document}